\documentclass{INTERSPEECH2023}

\interspeechcameraready 
\usepackage{booktabs}
\usepackage{multirow}
\usepackage{hyperref}

\title{A multimodal prototypical approach for unsupervised sound classification}

\name{Saksham Singh Kushwaha$^{1,2}$, Magdalena Fuentes$^{2,3}$}

\address{
  $^1$Courant Institute of Mathematical Sciences, New York University, NY, USA\\
  $^2$MARL, New York University, NY, USA\\
  $^3$IDM, New York University, NY, USA}
\email{sk8974@nyu.edu, mf3734@nyu.edu}

\begin{document}

\maketitle
 
\begin{abstract}
In the context of environmental sound classification, the adaptability of systems is key: which sound classes are interesting depends on the context and the user's needs. Recent advances in text-to-audio retrieval allow for zero-shot audio classification, but performance compared to supervised models remains limited. This work proposes a multimodal prototypical approach that exploits local audio-text embeddings to provide more relevant answers to audio queries, augmenting the adaptability of sound detection in the wild. We do this by first using text to query a nearby community of audio embeddings that best characterize each query sound, and select the group's centroids as our prototypes. Second, we compare unseen audio to these prototypes for classification. We perform multiple ablation studies to understand the impact of the embedding models and prompts. Our unsupervised approach improves upon the zero-shot state-of-the-art in three sound recognition benchmarks by an average of 12\%.

\end{abstract}

\section{Introduction}

Environmental sound event classification has several applications of interest to public health and industry such as assistive devices \cite{jain2022protosound}, autonomous navigation \cite{furletov2021auditory}, home assistants \cite{haeb2019speech}, noise mitigation \cite{bello2019sonyc}, among others. Typical sound recognition systems consist of deep-learning-based supervised models, where human annotations are needed to train a model to recognize sounds from a predefined set of classes. The main disadvantage of such models in practice is that they are very inflexible to work with out-of-domain sounds. Recent work has highlighted the importance of adaptability of sound recognition systems in the context of assistive devices \cite{jain2022protosound}, but this also holds in general: the vocabulary of sounds that such a system should recognize will vary from home to home, city to city, and application to application. It is troublesome to re-train such models each time. 

Many efforts have been made in making sound recognition models more adaptable, e.g. using few-shot learning \cite{wang2020few, wang2022hybrid} where only a few curated examples from each sound class are needed to train a competent model for environmental sound recognition and music classification \cite{wang2020few}. In the context of assistive devices, prototypical approaches have shown promise \cite{jain2022protosound}, also needing a few inputs from the user to select audios from a database or record them themselves. Although both approaches propose a useful and flexible change of paradigm with respect to previous supervised approaches, there is still the need for human supervision, and in cases where several sound classes want to be recognized, the time spent curating or selecting audio examples can be considerable. 

Recently, with the introduction of multimodal deep learning text and audio self-supervised models, the prospect of successfully doing zero-shot classification (classifying instances of unseen data without any training or fine-tuning) has improved considerably \cite{guzhov2022audioclip,elizalde2022clap}. This opens the possibility for domain adaptation of sound recognition systems by exploiting the correspondence of text and audio without any human intervention. 

In this work, we propose an unsupervised multimodal prototypical approach that leverages zero-shot text-to-audio retrieval capabilities of large multimodal models. To do so, unlike previous approaches, we use text embeddings to find representative audio clusters in the joint audio-text embedding space without any human supervision and compute the cluster's centroid as the prototype. At classification time, we use these audio prototypes to compare the unseen audio query and classify it. Our approach improves upon the zero-shot state-of-the-art in three well-known environmental sound classification benchmarks, namely ESC-50, UrbanSound8K, and FSD50k, and performs competitively to supervised approaches in a challenging multi-class scenario. Our contributions are as follows: 
1) we propose an unsupervised multimodal strategy to select audio prototypes using text for sound classification; 2) we evaluate the effectiveness of this approach using different datasets (single-label and multi-label) and different pre-trained text-audio models; 3) and we investigate the impact of prompting as well as cluster's size in the accuracy of our approach. Our code is open-source and available for research.
\footnote{ \scriptsize {\url{https://github.com/sakshamsingh1/audio_text_proto}}}

\section{Related work}
\label{sec:rel_work}

\begin{figure*}[ht!]
   \centering
   \includegraphics[width=\linewidth]{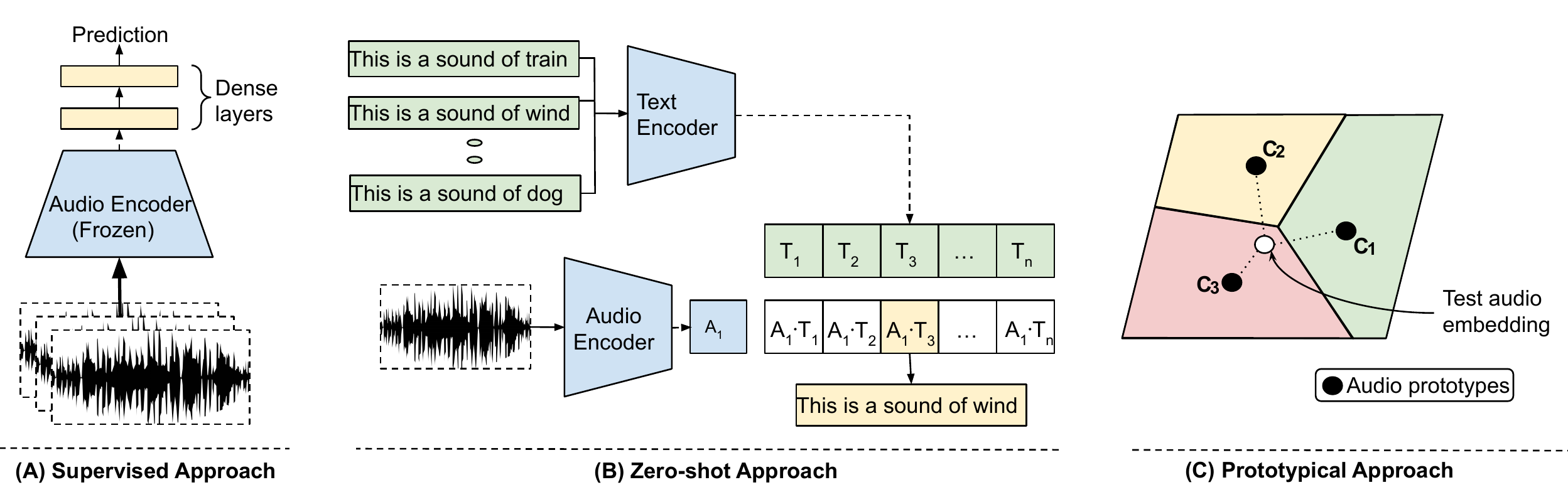}
   \caption{Typical classification approaches in the literature of environmental sound classification.}
   \label{fig:approaches}
\end{figure*}

\noindent \textbf{Supervised models}. 
Supervised models for environmental sound classification have recently shifted to rely heavily on transfer learning, the most popular approach being to pre-train audio-visual deep learning models using self-supervision \cite{cramer2019look,cartwright2019tricycle,Arandjelovic_2018_ECCV} on large amounts of data so it learns meaningful features from audio and images, and then using its audio encoder as input to a shallow classifier to work on new unseen audio data (see Figure \ref{fig:approaches}a). This is typically done by exploiting the semantically related information between the two modalities, typically audio and image, to algorithmically generate labels for large amounts of data and pre-train a large model without any human supervision. The fact that labels are generated automatically allows for exposing these self-supervised models to large amounts of data which would be impossible otherwise, and thus the superiority of these models with respect to supervised ones trained on small datasets. This transfer learning approach has proved to be effective in environmental sound classification \cite{cramer2019look, chen2022beats}, domestic sound classification \cite{politis2020overview}, among others. However, still relies on annotated labels for the fine-tuning stage, so the human intervention bottleneck remains. We include supervised models that leverage transfer learning from audio and text in our experiments for comparison to our prototypical approach, as explained in Section \ref{sec:method}.


\noindent \textbf{Text-audio deep learning models.} 
After the introduction and open release of CLIP \cite{radford2021learning}, a large text-image multimodal deep learning model which showed impressive results for text-image retrieval, zero-shot classification, image captioning\cite{mokady2021clipcap}, and text-to-image generation \cite{ ramesh2022hierarchical}, many approaches have been proposed to create models that have equivalent text-audio capabilities. Some of those approaches directly build on CLIP, e.g. by using its frozen image encoder to guide the training of an audio encoder that would learn embeddings in the pre-existing text-image joint embedding space \cite{wu2022wav2clip}. Other approaches fine-tune the text and image encoders along with an audio encoder in datasets that contain text, audio and image samples \cite{guzhov2022audioclip}. Other approaches train audio and text encoders from scratch using contrastive loss from audio and text pairs \cite{elizalde2022clap}, and even explore text-augmentation techniques to make the model more flexible to natural language inputs \cite{wu2022large}. As mentioned before, these text-audio models have shown great potential for zero-shot classification in new, unseen datasets, which is achieved by embedding audio samples and text labels into the same space and computing the similarity between them (see Figure \ref{fig:approaches}b). The biggest advantage of this approach is that it is completely unsupervised, but its main disadvantages are that it is sensitive to the ``quality'' of the prompt and its performance is still considerably lower than supervised approaches.  In this work, we leverage the potential of these models for zero-shot classification within a different approach: prototypical classification. For this, we explore the effectiveness of different pre-computed text-audio embeddings, in particular \cite{guzhov2022audioclip,wu2022large}, as they are the state-of-the-art in zero-shot environmental sound classification.

\noindent \textbf{Prototypical approaches.} 
Prototypical approaches have been successfully used in the context of computer vision \cite{snell2017prototypical} and sound recognition \cite{jain2022protosound}. These approaches typically consist of a first stage of selecting a small set of examples that are characteristic of a class, and then obtain the centroid of each class group as the prototype. Either if the embeddings are pre-trained \cite{jain2022protosound} or learned in the process \cite{snell2017prototypical} of computing clusters and centroids, similarly to few-shot learning, prototypical approaches typically rely on few annotated data (the examples). We propose to select the examples without any human intervention, by leveraging the text-to-audio zero-shot capabilities of multimodal deep learning models. We do this by converting both text and audio into embeddings and using the proximity of those embeddings to query audio using text. What this means in practice is that the user would input a text prompt as query, and the model would internally retrieve a relevant audio prototype to represent the user query or label, without the need of further recording or choosing between examples on the users' side. Our method is explained in detail below.

\section{Method}
\label{sec:method}

\begin{figure*}[ht!]
   \centering
   \includegraphics[width=\linewidth]{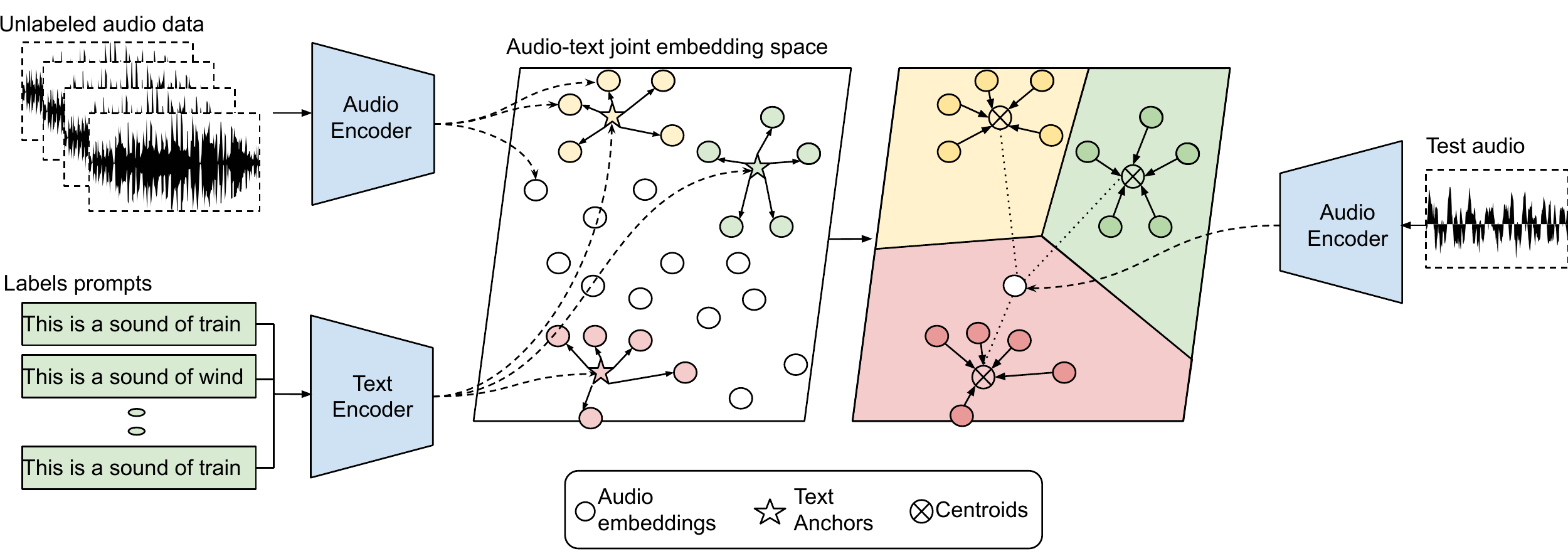}
   \caption{
   Our unsupervised multimodal prototypical approach. We use text queries as anchors to retrieve clusters of audio embeddings that best represent the text. The centroids of these clusters serve as prototypes. During inference, the new audio is compared to the prototypes and assigned the corresponding text label.
   }
   \label{fig:models}
\end{figure*}

\subsection{Datasets and metrics}
We use three audio classification datasets for our experiments that differ in size, number and type of labels. These datasets are described below.

    \noindent \textbf{ESC-50}\cite{piczak2015esc}: The ESC-50 dataset compromises of 2000 environmental audio recordings, with each clip of 5 seconds. The audio clips belong to 50 class labels that can be divided into 5 major categories such as animals and urban noises. The dataset is divided into 5 non-overlapping folds by the authors for cross-validation. The models are evaluated using 5-fold multiclass classification accuracy.
    
    \noindent \textbf{UrbanSound8K}(US8K)\cite{salamon2014dataset}: This dataset consists of 8732 recordings (each track $\le$ 4s) which belong to 10 categories (eg. car horn, children playing). Similar to ESC-50, this dataset is also divided into 10 non-overlapping folds and is evaluated using 10-fold multiclass classification accuracy.
    
    \noindent \textbf{FSD50K}\cite{fonseca2021fsd50k}: This dataset consists of 51,197 Freesound\cite{fonseca2017freesound} that span over 200 classes. The clips have varying lengths ranging from 0.3s to 30s and are organized hierarchically (144 leaf nodes and 56 intermediate nodes) with a subset of the AudioSet Ontology. The dataset is a multi-label dataset and has been divided into train, validation, and test split. To evaluate the performance of models trained on this dataset, the mean average precision(mAP) metric has been adopted.

\subsection{Multimodal prototypical approach}
\label{ssec:proto} 
Our approach is illustrated in Figure \ref{fig:models}. It consists of two main steps: 1) retrieving audio prototypes from text, and 2) using these prototypes for classification. Additionally, we explore the impact of prototype selection, as explained below.

\noindent \textbf{Prompt selection.} The performance of text-audio models for zero-shot classification is sensitive to the particular text prompts used to query \cite{elizalde2022clap}, given the data that was used to train them \cite{wu2022large}. To analyze that and mitigate its impact in our study, we use different formulations of prompts that include the labels of each dataset and compare the performance of the different models for the ESC-50 dataset. Based on the accuracy performance of each configuration, we select the best prompt for each model and use it for the remaining experiments. In practice, this can be done in an annotated dataset different from the target data. 

\noindent \textbf{Embedding models.}
Our prototypical approach leverages pretrained audio and text encoders from two state-of-the-art multimodal models, namely AudioClip \cite{guzhov2022audioclip} and LAION-CLAP \cite{wu2022large}. Specifically, we refer to our approach based on 
AudioClip as Proto-AC. Additionally, our approach that employs the encoders of LAION-CLAP with keyword-to-caption and feature fusion is referred as Proto-LC.

\noindent \textbf{Unsupervised selection of audio prototypes using text.}
As depicted in Figure \ref{fig:models}, our approach uses text queries (represented by labels prompts from each dataset) as anchors in the joint audio-text embedding space to retrieve local neighborhoods (clusters) of audio embedding that better represent the text query. We use $k$-nearest neighbors for this, where $k$ was chosen via a grid search for ESC-50 but kept the same for the rest of the datasets. The reasoning behind this is that we want to strike a reasonable number for $k$ but we do not want to tune it for each dataset since this would require using the labels in practice. The centroids of those clusters become the  prototypes for the different sound classes of interest. As stated before, our interest in exploiting the retrieval capabilities of these multimodal models is to have a completely unsupervised approach, that intuitively will have better audio-text matches than zero-shot learning given that the matching process is done considering multiple examples (the clusters) instead of one, and similarities are measured between embeddings of the same modality to make the final assignments. 

\noindent \textbf{Classifying unseen audio samples.}
Given a new audio sample to be classified, we first extract its embedding using the same audio encoder that is used for computing the clusters and prototypes. Then we proceed in two different ways depending if the dataset is single-label or multi-label. 
For single-label datasets like ESC-50 and US8K, we choose the predicted label to be the one whose prototype is closest (via cosine similarity) to the embedding of the unseen audio. In the case of multi-label datasets like FSD50K, we compute a vector with the different classes' likelihoods by computing the sigmoid of the cosine-similarity between the embedding of the unlabeled audio and the embedding of each of the prototypes. We then calculate the mAP with multi-label targets as one-hot vectors.

\subsection{Comparison to other approaches}

\noindent \textbf{Zero-shot classification.} 
We use the same pre-trained embedding models as our prototypical approach to do zero-shot classification as a baseline. To do so, we first compute the embeddings for all the test audio and prompted text labels using the pretrained encoders. Because text and audio share the same embedding space we compute cosine similarity between their embeddings.
We then use softmax and sigmoid over this distribution for single-label and multi-label classification respectively.

\noindent \textbf{Supervised classification with pre-trained embeddings.} Using the same embedding models as before (LAION-CLAP and AudioCLIP), we follow the supervised approaches explained in Section \ref{sec:rel_work}, pre-compute the audio embeddings and use them as input to a shallow classifier. 
The classifier consists of 3 fully connected layers with relu activations in all of them except the last. For the last layer, in the case of single-label classification we use a softmax activation, and for multi-label classification we use a sigmoid activation. We train this network using Adam with learning rate of 1e-4, $\beta_1=0.9$ and $\beta_2=0.999$.

\noindent \textbf{Supervised prototypical networks.} To understand how good are the embeddings to characterize each sound class within each dataset, we include two baselines (one with LAION-CLAP and another with AudioClip) in which we select the audio clusters using their labels directly. We then compute the centroids as prototypes and perform inference exactly as the prototypical approaches explained before. 

\label{ssec:sup}

\begin{table*}[htb]
\centering
\resizebox{\textwidth}{!}{%
\begin{tabular}{lcccccc}\toprule
                & \multicolumn{2}{c}{ESC-50 (acc)} & \multicolumn{2}{c}{US8k (acc)} & \multicolumn{2}{c}{FSD50K (mAP)} \\
\cmidrule(lr){2-3} \cmidrule(lr){4-5} \cmidrule(lr){6-7} 

                & Zero Shot     & Supervised    & Zero Shot      & Supervised    & Zero Shot   & Supervised   \\
\midrule                
Wav2Clip        & 0.41          & 0.86          & 0.40           & 0.81          & 0.03          & 0.43       \\
AudioClip\textsuperscript{*,\textdagger}      & 0.68          & 0.88          & 0.62           & 0.86          & 0.20          & 0.50          \\
CLAP            & 0.83          & \textbf{0.97} &\textbf{ 0.73}  & 0.88          & 0.30          & 0.59       \\
LAION-CLAP\textsuperscript{*,\textdagger}      & 0.91          & 0.96          & 0.72           & \textbf{0.89} & 0.22          & 0.61         \\
\midrule
Proto-AC\textsuperscript{\textdagger}      & 0.78          & 0.82          & 0.71           & 0.77          & 0.40          & 0.48         \\
Proto-LC\textsuperscript{\textdagger}      & \textbf{0.96} & \textbf{0.97} & \textbf{0.73 } & 0.83          & \textbf{0.52} & \textbf{0.65}         \\ \bottomrule
\end{tabular}
}
\caption{Classification results for the different approaches and configurations. $^*$Results reproduced using author's code
\textsuperscript{\textdagger}Best label prompts were selected based on the performance on ESC-50}
\label{tab:results}
\end{table*}

Our results are presented in Table \ref{tab:results}. We group results as \textit{zero-shot} when methods do not use any label, and \textit{supervised} when the labels are used as explained in Section \ref{ssec:sup}. A first observation is that the prototypical approach performs better in most cases, for all datasets, with the best configuration being Proto-LC. In the following, we break down the discussion in the different aspects of this study.

\section{Results and discussion}

\noindent \textbf{What is the effect of the prompt?} We analyze the effect of the prompt by trying different prompting variations as text anchors, as shown in Table \ref{tab:prompts}. We examine the performance of the embedding models as well as the prototypical models in zero-shot in the ESC-50 dataset. We selected 5 prompts that show promise in previous works \cite{wu2022large}. A first observation is that the different embedding models have different robustness to changing the prompts. AudioClip's performance variation is relatively small (2\% maximum), while LAION-CLAP's variation is larger with up to 9\% difference in performance. Surprisingly, that trend does not transfer to the prototypical models that use AudioClip and LAION-CLAP respectively: the variation for both Proto-AC and Proto-LC is relatively large (8\% and 4\% respectively), with AudioClip having the least variability. 

\begin{table}[htb]
\centering
\resizebox{\columnwidth}{!}{%
\begin{tabular}{lcccc}
\toprule                                                                             
& \multicolumn{1}{c}{AudioClip} & \multicolumn{1}{c}{LAION-CLAP} & \multicolumn{1}{c}{Proto-AC} & Proto-LC \\ 
\cmidrule(lr){2-5} 
Prompt 1                     & 0.67          & 0.83          & 0.77          & 0.94          \\
Prompt 2          & 0.68          & 0.86          & 0.72          & 0.92           \\
Prompt 3 & \textbf{0.69} & 0.90          & 0.72          & \textbf{0.96}   \\
Prompt 4             & 0.68          & 0.88          & \textbf{0.78} & 0.95             \\
Prompt 5  & 0.67          & \textbf{0.92} & 0.70          & \textbf{0.96}      \\ \bottomrule
\end{tabular}%
}
\caption{Accuracy on ESC-50 with different prompts. Prompt 1: `\{Class label\}', Prompt 2: `I can hear \{class label\}', Prompt 3: `This is an audio of \{class label\}', Prompt 4: `This is \{class label\}', Prompt 5: `This is a sound of \{class label\}'.}
\label{tab:prompts}
\end{table}
\vspace{-0.5cm}

\noindent Another surprising finding not included in Table \ref{tab:prompts} was 
that the models' performance changed if the prompts started with an uppercase or lowercase letter, where the lowercase configuration performed considerably worse (an average of 5\%). This shows how sensitive this models are to prompting, and indicates that further augmentations beyond rephrasing are needed during training to ensure robustness in practice. Based on the results in Table \ref{tab:prompts} we chose the best prompt for each model according to this ablation study for the remaining experiments.

\begin{figure}[h!]
  \centering
\includegraphics[width=0.63\linewidth]{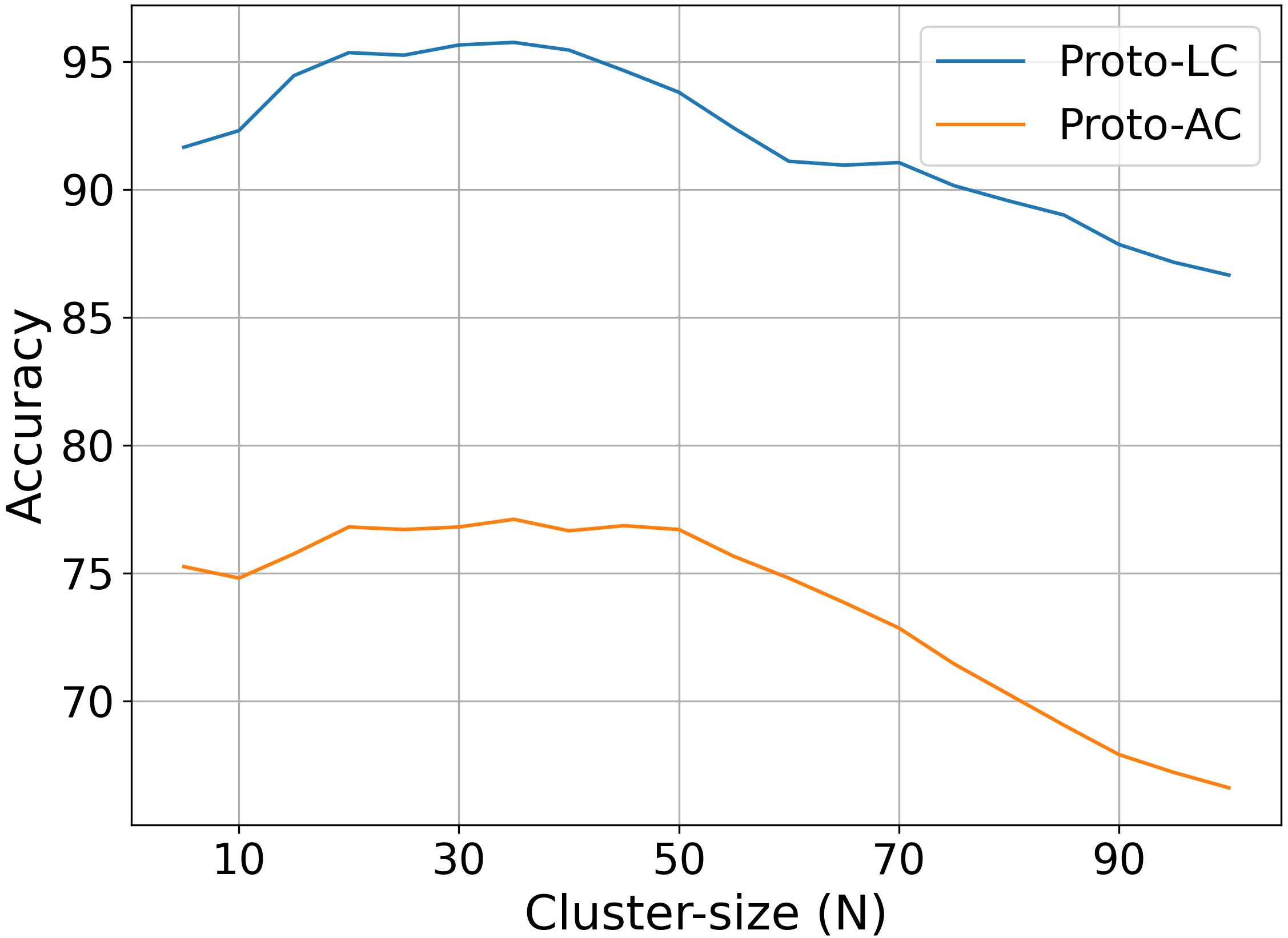}
  \caption{Cluster size vs accuracy of prototypical methods.}
  \label{fig:acc}
\end{figure}

\noindent \textbf{The impact of the number of neighbours.}
In order to obtain the clusters of audio examples, we need to choose how many examples we consider in each cluster. Intuitively, this hyper-parameter ($k$) will impact the quality of the prototypes, since too few or too many examples could lead to an inaccurate prototype choice. The selection of such a hyper-parameter is the only stage in the prototypical approach that would require the use of labels, if we want to compute the performance of the model at different values of $k$ and choose the optimal one. As explained in Section \ref{ssec:proto}, we perform a preliminary study in ESC-50 to understand the impact of such parameter, strike a reasonable value for $k$ and keep the same value to the other datasets to simulate how the method would be used in practice without this information. Figure \ref{fig:acc} shows this for the ESC-50 dataset. As shown there, the performance of the model is equally high for a large set of values of $k$. We choose $k=35$, and \textit{we keep this value for the rest of the datasets}. 
As shown in Table \ref{tab:results}, despite not optimizing $k$ for US8K or FSD50K, the prototypical approach outperforms the other methods, showing promise in robustness and low parameter tuning.

\noindent \textbf{Performance in single-label vs. multi-label datasets.}
Table \ref{tab:results} shows that the prototypical approach is significantly better than the zero-shot baseline, especially in the FSD50K dataset compared to US8K or ESC-50. This implies that Proto-AC/Proto-LC are more effective in handling complex relationships in multi-label datasets than the zero-shot method. The reason for this may be that computing the prototypes as centroids of nearby audios leads to a better alignment of the prototypes with the audio embeddings than the text embeddings, which we will further investigate in future work.

\section{Conclusions}
Our work proposes using text-audio multimodal deep learning model capabilities to classify environmental sounds using prototypical classification, without the need for human intervention. Our method performs better than zero-shot classification and can enable user-adaptable sound recognition systems through text. For future research, we will investigate training encoders to be more robust to prompt changes, as well as compare the computational complexity of different approaches.

\bibliographystyle{IEEEtran}
\bibliography{mybib}

\end{document}